\documentstyle[12pt,epsfig]{article}
\newcommand {\be}{\begin{equation}}
\newcommand {\ee}{\end{equation}}
\newcommand {\ba}{\begin{eqnarray}}
\newcommand {\ea}{\end{eqnarray}}

\begin{document}

\def \a'{\alpha'}
\baselineskip 0.65 cm
\begin{flushright}
IPM/P-2006/033\\
\today
\end{flushright}
\begin{center}{\large
{\bf Reconciling large  CP-violating phases with  bounds
on the electric dipole  moments in the MSSM }}
 {\vskip 0.5 cm}

{\bf  Seyed Yaser Ayazi   and Yasaman Farzan
}{\vskip 0.5 cm }
Institute for Studies in Theoretical Physics and Mathematics (IPM)\\
P.O. Box 19395-5531, Tehran, Iran\\
\end{center}

\begin{abstract}
The possibility of cancelation between different contributions to
$d_e$, $d_n$ and $d_{Hg}$ has been reconsidered with special
emphasis on the region that is phenomenologically interesting
(intermediate values of $\tan \beta$ and sub-TeV sfermion masses).
It is found that in the range favored by electroweak baryogenesis
({\it i.e.,} $|\mu|\simeq M_1$ or $|\mu|\simeq M_2$), $\sin
[\theta_\mu+\theta_{M_1}]\sim 1$  can be compatible with the EDM
bounds even for slepton masses below 500~GeV. Such large values of
the phases promise a successful electroweak baryogenesis. The
possibility of large CP-odd effects at linear collider has also
been discussed.
\end{abstract}
\section{Introduction}
Elementary particles can possess Electric Dipole Moments (EDMs),
only if the CP symmetry is violated. For this reason, studying
EDMs of the elementary particles is of prime importance as it can
teach us about CP-violation which is closely related to the
creation of the baryon asymmetry of the universe. It is well-known
that within the Standard Model (SM) of the elementary particles
 violation of CP can take place. In fact in the Kaon and B-meson sector, the CP
symmetry has been observed to be violated in accordance with the
prediction of the Standard Model.
However, the maximum possible values of EDMs
in the context of SM are extremely small; the SM predicts $d_e\sim
10^{-38}\ e \ {\rm cm}$ \cite{e} and the prediction of SM for
$d_n$ ranges from $ 10^{-31}\ e \ {\rm cm}$  to $ 10^{-33}\ e \
{\rm cm}$ \cite{n}. So far no electric dipole moment for the
electron or neutron has been detected but strong bounds on these
quantities have been obtained \cite{pdg,commins,cerncourier}
\begin{equation}
|d_e|<1.4\times 10^{-27} \ e \ {\rm cm}
 \ \ \ \ \ \ |d_n|<3.0 \times 10^{-26} \ e \ {\rm cm}.\label{10-27}
 \end{equation}
There are good prospects of improving these bounds  by several
orders of magnitude in the next few years \cite{prospect}. Values
of EDMs much larger than the SM prediction would indicate new
sources of CP-violation with origin in physics beyond the SM.

The Minimal Supersymmetric Standard Model (MSSM) is arguably the
most popular model beyond the SM. The general  MSSM introduces 44
sources of CP-violation \cite{godbole}. Mainly for the sake of
simplicity, studies in the literature are concentrated  on the
mSUGRA model (which is also called constrained MSSM or cMSSM)
which assumes that at the GUT scale the masses of scalar
components of the chiral superfields are unified to $m_0$ while
the masses of gauginos are also unified to $m_{1/2}$. In this
model, the A-terms (the trilinear scalar terms in the soft
supersymmetry breaking potential) are also universal and set to be
proportional to the corresponding Yukawa couplings: at the GUT
scale, $A_\ell=a_0 Y_\ell, \ A_u =a_0 Y_u \ {\rm and} \ A_d=a_0
Y_d $. In the constrained MSSM the number of independent
CP-violating phases are reduced to two which  are usually
attributed to the phases of $a_0$ and the mu-term (bilinear Higgs
mass term in the superpotential). Taking the values of the
parameters at their phenomenologically favorable range
($m_{1/2}\sim m_0 \sim 100 \ {\rm GeV}, \ \tan \beta\sim 10, \
\theta_\mu\sim 1 \  {\rm and } \ \theta_{A_e} \sim 1$), one finds
that the EDMs of the electron, neutron and mercury exceed the
experimental bounds  by several orders of magnitude. In principle,
to suppress the EDMs to below their experimental bounds, three
possibilities exist:
\begin{itemize}
\item The  first generation of sleptons and the first two generation of squarks are very heavy \cite{heavymasses}.
This means the production and study of these particles at ILC and LHC will be difficult, if possible
at all.
 The other reason that this
possibility is in disfavor is that, with large sfermion masses the
annihilation rate of the Lightest Supersymmetric Particle (LSP)
may be too low and as a result the relic density of the LSP may be
larger than the observed dark matter density.

\item The phases of $\mu$ and $a_0$ are both zero or very small which
means that there will not be any interesting display of
CP-violation in colliders. Moreover, electroweak  baryogenesis
cannot take place in this case \cite{rubakov,stefano}.
\item
The contributions of the phases of $\mu$ and $a_0$ cancel each
other. From the phenomenological point of view, this is the most
interesting solution because it leaves room for a host of
non-trivial CP-violating as well as CP-conserving phenomena to be
discovered.
\end{itemize}

The third  possibility has been extensively studied in the
literature \cite{cancelation} and  unfortunately it seems that
cancelation scenario works only if the phase of $\mu$ is ${\cal
O}(10^{-2})$ or less which is too small to result in detectable
CP-violating effects in colliders. This is due to two  reasons: 1)
The severe upper bounds on $d_{Hg}$ and $d_n$ have to be
simultaneously satisfied that is while  there are only two
CP-violating phases. It seems natural that in the parameter range
that the contributions of $\theta_{a_0}$ and $\theta_{\mu}$ to
$d_e$ cancel each other, their contribution to $d_{Hg}$ add up and
vice versa. 2) In the large $\tan \beta$ regime (which is favored
by the LEPII data \cite{pdg} as well as the SO(10) model
\cite{so(10)}), the contribution of $\theta_\mu$ to the EDMs of
the electron as well as the  down quark is enhanced such that it
cannot be canceled by the effect of the phase of $a_0$, unless the
phase of $\mu$ itself is small. This means that, for large
$\tan\beta$ regime, even if we relax the condition of universality
of the $A$-terms at the GUT scale (taking $A_e$, $A_u$ and $A_d$,
at the low energy scale, arbitrary), cancelation condition can be
hardly satisfied \cite{olive1}.

 In this paper, we  relax  some of
the universality conditions and find that for a range of
parameters, which from phenomenological point of view is
interesting, the cancelation scenario is revived even for
intermediate values of $\tan \beta$ ($\tan \beta\simeq 10$). This
basically happens when the masses of sfermions are low while the
values of $M_2$ (Wino mass), $A_e$ and $A_d$ are large. Putting it
in another way, the suppression scenario discussed here engages
two of the above conditions: cancelation as well as having large
mass parameters. However, we only take some of the parameters
large; i.e., there is a small hierarchy of order of 10 among the
supersymmetric parameters at the electroweak scale. Such a
hierarchy is not by any means peculiar to this model. Even in the
context of the mSUGRA model, at the low scale the colored
particles are expected to be (5$-$10) times heavier than those
which are singlets of SU(3).

In section 2, we describe the model and discuss that having $A_e$
and $A_d$ at the TeV scale can be consistent with the bounds from
the Color and Charge Breaking (CCB) vacua considerations. In
section 3, we study the behavior of EDMs by varying different
parameters and discuss the robustness of our results. In section
4, we study the possibility of cancelation in the parameter range
favored by resonant electroweak baryogenesis. In section 5, we
discuss if the cancelation opens the possibility of large enough
phases  to cause sizeable CP-violating effects in colliders.
Conclusions are summarized in section 6.
\section{The model} In this paper, we
consider the Minimal Supersymmetric Standard Model with
superpotential
 \be
   W_{MSSM} =  Y_{u}\widehat{{u}^c}  \ \widehat{Q} \cdot \widehat{H_{u}}
        -Y_{d} \widehat{{d}^c} \ \widehat{Q} \cdot \widehat{H_{d}}
        - Y_{e}  \widehat{{e}^c} \ \widehat{L}  \cdot \widehat{H_{d}}
-\mu\ \widehat{H_{u}}\cdot \widehat{H_{d}} \ee
 where, $\widehat{L}$, $\widehat{Q}$, $\widehat{H_{u}}$, $ \widehat{H_{d}}$
 are doublets of chiral superfields associated respectively with
 left-handed leptons, left-handed quarks and the two Higgs doublets of the
 MSSM. In the above formula,
 $\widehat{{u}^c}$, $\widehat{{d}^c}$ and $\widehat{{e}^c}$ are
the chiral superfields associated with the corresponding
right-handed fields. The  soft supersymmetry breaking part of
Lagrangian, at the electroweak scale, is taken to have the form
\ba \label{MSSMsoft}\L_{\rm soft}^{\rm MSSM} &=&-\ 1/2 \left( M_3
\widetilde{g} \widetilde{g} + M_2 \widetilde{W} \widetilde{W}+ M_1
\widetilde{B}\widetilde{B} +{\rm  H.c.} \right) \cr &-&(A_{u
i}Y_{u ii} \widetilde{{u_{i}}^c} \ \widetilde{Q_{i}} \cdot
H_{u}-A_{d i}Y_{d ii} \widetilde{{d_{i}}^c} \ \widetilde{Q_{i}}
\cdot H_{d}-A_{e i}Y_{e ii} \widetilde{{e_{i}}^c} \
\widetilde{L_{i}} \cdot H_{u} + {\rm H.c.} )\cr  &-& \widetilde{Q
_{i}}^{\dag} \ m_{\tilde{Q} ii}^{2}\widetilde{Q_{i}} -
\widetilde{L_{i}} ^{\dag} \ m_{\tilde{L} ii}^{2}\widetilde{L_{i}}
- \widetilde{(u_{i}^c)} ^{\dag} \ m_{\tilde{u}
ii}^{2}\widetilde{u_{i}^c} - \widetilde{(d_{i}^c)} ^{\dag} \
m_{\tilde{d} ii}^{2}\widetilde{d_{i}^c} - \widetilde{e_{i}^c
}^{\dag} \ m_{\tilde{e} ii}^{2}\widetilde{e_{i}^c}\cr &-& \
m_{H_{u}}^{2}\ H_{u}^{\dag}\ H_{u}-\ m_{H_{d}}^{2}\ H_{d}^{\dag}\
H_{d}-(\ B_H \ H_{u}\cdot H_{d}+ {\rm H.c.}),\ea where the ``$i$"
indices determine the flavor. We have relaxed the universality
assumption ({\it i. e.,} in general, $m_{\tilde{\mu}}^2\ne
m_{\tilde{e}}^2 \ne m_{H_u}^2 \ne {\rm etc}$); however, we have
taken a flavor conserving soft potential ({\it i.e.,} there is no
off-diagonal terms in the flavor basis).  The latter assumptions
is  motivated by observation \cite{flavor}. We have defined the
$A$-parameters  factoring out the corresponding Yukawa couplings.
Notice that since we do not assume any universality, from the
beginning we concentrate on the potential at the electroweak scale
rather than some high GUT scale.
 Hermiticity of the Lagrangian implies that  $m_{H_u}^2$,
$m_{H_d}^2$  and  the  sfermion masses  are
real.
 The rest of parameters in Eq.~\ref{MSSMsoft} can in general be complex.
 By rephasing the
fields we can make $B_H $ and $M_2$ real but, in this basis  the
phases of $\mu$- and $A$-parameters as well as $M_3$  and $M_1$
cannot be rotated away and should be considered as new sources of
CP-violation.

In Ref \cite{plehn}, the possibility of cancelation between
contributions of the phases of $A_e$, $A_d$, $A_u$, $M_1$, $M_3$
and $\mu$ has been studied for $|A_i|<1$~TeV and $\tan \beta<10$.
As expected, they have found that increasing $\tan \beta$, the
range of $\theta_\mu$ for which cancelation is possible shrinks.
In this paper, we focus on $|A_s|, |A_d|>1$~TeV and show that, for
large values of $|A_i|$, cancelation scenario is revived even for
the intermediate values of $\tan \beta$.

One can reconstruct models for which deviation from universality
is partial. For example, it is shown \cite{yasaman} that in the
context of type I seesaw mechanism embedded in the MSSM, the
neutrino B-term radiatively induces different corrections to $A_u$
and $A_e$ (lifting their universality) however, in this model, the
universality of gaugino masses (at the GUT scale) is maintained.
Considering such possibilities, we first perform an analysis
setting the phases of gaugino masses  equal to zero and then allow
the phases of $M_1$ and $M_3$ to be nonzero. Of course, allowing
more phases to be nonzero, the likelihood of cancelation
increases.

 For large values
of $A$-terms, one of course has to check the CCB bounds. In the
following, we discuss that by relaxing the condition of
universality at the GUT scale, we can have values of $A_e$ and
$A_d$ as large as a
few TeV while keeping the sfermion masses below TeV
without encountering Color or Charge Breaking (CCB) vacua. In the
end, we suggest a way to test the preassumption that goes into
this possibility.

 As it is well-known, at the tree level, the conditions for the electroweak
symmetry breaking are \be |\mu|^2+m_{H_d}^2=B_H  \tan
\beta-\frac{m_Z^2}{2}\cos 2 \beta \label{Mhd}\ee and
$$ |\mu|^2+m_{H_u}^2=B_H  \cot \beta+\frac{m_Z^2}{2}\cos 2 \beta.$$
For large $\tan \beta$, we can neglect $B_H  \cot \beta$ and write
\ba m_{H_u}^2&\simeq&-|\mu|^2+\frac{m_Z^2}{2} \cos 2 \beta\cr
m_{H_d}^2&\simeq&B_H  \tan \beta+m_{H_u}^2-m_Z^2 \cos 2\beta ,\ea
 so we expect
$m_{H_u}^2$ to be negative.  In the mSUGRA, the values of
$m_{H_u}^2$, $m_{H_d}^2$ as well as the sfermion and gaugino
masses are all determined by two parameters $m_0$ and $m_{1/2}$.
This means that to have a low spectrum, $B_H \tan \beta$ should be
small, too. For large values of $\tan \beta$, this means that $B_H
$ should  be much smaller than other supersymmetric parameters
which sounds unnatural (see however \cite{however}). Now if we
relax the unification of the masses at high energies and take $B_H
$ to be of order of $|m_{H_u}^2|$, we find that $m_{H_d}^2$ can be
positive and large.

For positive $m_{H_d}^2$ as discussed in \cite{gunion} to
guarantee that no CCB occurs, it is sufficient to have \footnote{
The bounds (\ref{aebound},\ref{adbound}) are based on tree-level
analysis. However, \cite{casas} shows that loop effects do not
affect these results.} \be \label{aebound} A_e^2
<3(m_{H_d}^2+m_{\tilde{e}_L}^2+m_{\tilde{e}_R}^2)\ee and \be
\label{adbound} A_d^2
<3(m_{H_d}^2+m_{\tilde{d}_L}^2+m_{\tilde{d}_R}^2)\ee  The larger
$m_{H_d}^2$ and $B_H  /\cos \beta$, the less stringent the upper
bounds on $A_e$ and $A_d$. Remembering that the masses of CP-odd
Higgs field ($A^0$), the heavier CP-even Higgs ($H^0$) and charged
Higgses ($H^\pm$) are given by $(2 B_H/\sin 2 \beta)^{1/2}$, the
assumption of large $m_{H_d}^2$ can put into test at the LHC
\cite{hashemi}.
 Finally, since
we are assuming that off-diagonal elements of the $A$-terms are
absent, we do not need to be concerned about the region unbounded
from below \cite{casas}.

\section{Numerical results} In this section, we study the electric
dipole moments of the electron, mercury and neutron to find ranges
of phases for which cancelation is possible.
We first discuss the constraints on the input parameters from
various observations with special emphasis on the uncertainty in
the input parameters and theoretical formulae. We then discuss
how, by varying the input variables, the overall behavior of the
bounds changes. We then analyze the possibility of cancelation in
the following two cases that are phenomenologically interesting:
1) close to the benchmark SPS1a' \cite{sps1a'} ($\equiv$B''
\cite{b''}); 2) the range $\mu\simeq M_1$ where the Higgsino-Bino
mixing is large. The latter case is of interest because in this
regime, neutralino annihilation is relatively large, yielding the
desirable dark matter density.

We assign various values in the relevant range to CP-conserving
parameters of the model ($\tan \beta$, the $\mu$ parameters, the
$A$-parameters, the selectron masses, the masses of superpartners
of light quarks and gaugino masses). As it is well-known, the
condition for electroweak  symmetry breaking determines the values of
$\mu$ in terms of $m_{H_d}^2$, $m_{H_u}^2$ and $\tan \beta$. In
this paper, as we discussed earlier we do not make a priori any
assumption on the values of $m_{H_u}^2$ and $m_{H_d}^2$ so we are
free to assign any value to $|\mu|$. In this regard, our model
resembles the Non-Universal Higgs Mass (NUHM) model which  has
recently received attention in the literature \cite{nuhm}.

\begin{figure}
\psfig{figure=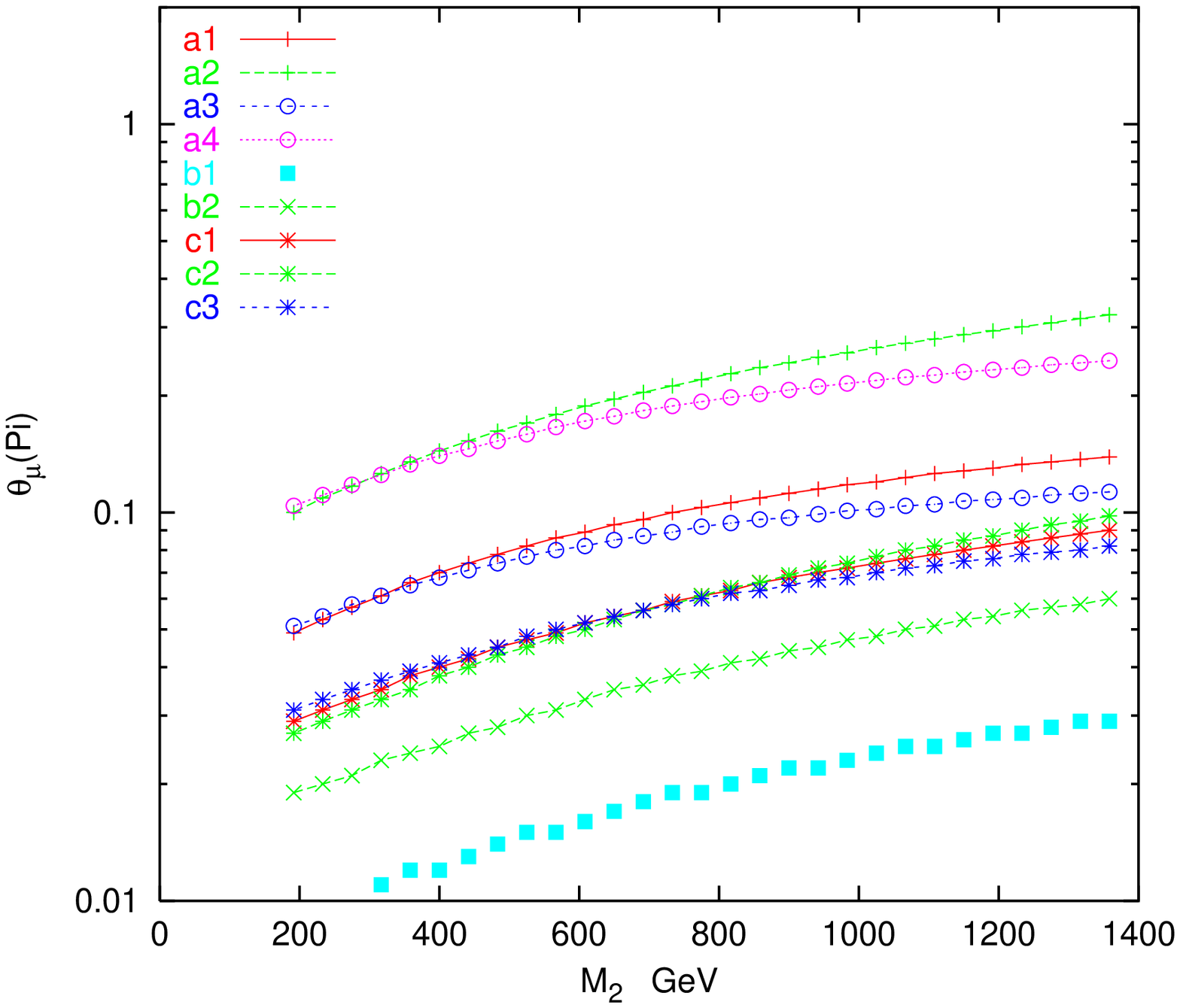,bb=33 37 553 501, clip=true, height=4
in} \caption{Maximum values of $\theta_\mu$ for which cancelation
between contributions of the phases of $A_e$ and $\mu$ to $d_e$ is
possible.   For curves (a1-a2-a3-a4), $\tan \beta=10$,
$|\mu|=440$~GeV, $m_{\tilde{e}_L}=305$~GeV,
$m_{\tilde{e}_R}=175$~GeV, $m_{\tilde{\nu}_L}=295$~GeV and
$M_1=155$~GeV. The values of $|A_e|$ and $|\mu|$ for these curves
are as follows: a1) $|A_e|=2000$~GeV and $|\mu|=440$~GeV; a2)
$|A_e|=4000$~GeV and $|\mu|=440$~GeV; a3) $|A_e|=2000$~GeV and
$|\mu|=550$~GeV; a4) $|A_e|=4000$~GeV and $|\mu|=550$~GeV. For
curves (b1) and (b2), $\tan \beta=20$, $|\mu|=500$~GeV,
$m_{\tilde{e}_L}=450$~GeV, $m_{\tilde{e}_R}=345$~GeV,
$m_{\tilde{\nu}_L}=440$~GeV and $M_1=175$~GeV. For (b1),
$|A_e|=2000$~GeV while for (b2) $|A_e|=4000$~GeV. For (c1-c2-c3),
$\tan \beta=10$, $m_{\tilde{e}_L}=405$~GeV,
$m_{\tilde{e}_R}=255$~GeV, $m_{\tilde{\nu}_L}=400$~GeV,
$M_1=195$~GeV and $|A_e|=2000$~GeV. The values of $|\mu|$ for
curves (c1), (c2) and (c3) correspond to $|\mu|=500$~GeV, $|\mu|=
400$~GeV and $|\mu|= 600$~GeV, respectively.} \label{deM2new}
\end{figure}

One of the triumphs of the MSSM is providing a natural candidate
for the dark matter; i.e. the lightest neutralino
($\tilde{\chi}_1^0$). In order for the relic density of
neutralinos to satisfy the precise results of WMAP, the parameter
space of the constrained MSSM is reduced to narrow strips in the
$m_0-m_{1/2}$ plane  for  given values of $\tan \beta$ and $A_0$
\cite{petra}. The contribution of neutralino to Dark matter energy
density is determined by annihilation rate after the temperature
drops below its mass. In the present model, it is  possible to
tune the (co)annihilation rate of neutralino to a value that
explains the data: For the parameter range that we have chosen
$\tilde{\chi}_1^0$ has a large  $\tilde{B}^0$ component which
means it has a large annihilation cross section with
$\tilde{\tau}_R$. If the mass of $\tilde{\tau}_R$ is close to that
of $\tilde{\chi}_1^0$, for temperatures slightly below their mass,
their density will be comparable. This means the coannihilation
rate of neutralinos with $\tilde{\tau}_R$ can be large enough to
bring the contribution of neutralinos to the dark matter density
to the desired values. Notice that in this scenario the dark
matter density is not sensitive to the value of $M_2$ (Wino mass)
because (i) the $\tilde{W}^0$ component of  $\tilde{\chi}_1^0$  is
small; (ii) $\tilde{\tau}_R$ does not couple to $\tilde{W}^0$. So,
varying the values of $M_2$ (as in Fig. \ref{deM2new}), will not
considerably affect the neutralino density.

  Another major
constraint on the MSSM parameters comes from the radiative
correction to Br$(b \to s \gamma)$ (see {\it e.g.,}
\cite{btosgamma}). The leading 1-loop SUSY diagrams involve loops
with a charged Higgs and a top quark and loops with a chargino and
a squark. Since in the present analysis the deviation from the
cMSSM spectrum is in the direction of increasing $M_2$ and $B_H $
(and hence $m_{\tilde{\chi}^+}$ and $m_{H^+}$) the SUSY correction
to $b \to s \gamma$ is further suppressed and as a result the
bounds from $b \to s \gamma$ cannot be significant.

 To
calculate the EDMs and CEDMs of the elementary particles, we use the formalism developed in \cite{nath}.
In the literature, there are various different formulae for the
EDM of mercury:
\begin{itemize}
\item according to \cite{falk}
 \be \label{Hgfalk}d_{Hg}=-(\tilde{d}_d-\tilde{d}_u -0.012
\tilde{d}_s)\times 3.2 \cdot 10^{-2} e, \ee where $\tilde{d}_d$,
$\tilde{d}_u$ and $\tilde{d}_s$ are respectively the
chromoelectric dipole moments of the $d$, $u$ and $s$ quarks.
\item that is while according to \cite{shimizu1}
\be \label{Hgshimizu}d_{Hg}=8.7\times 10^{-3} \times
e(\tilde{d}_d-\tilde{d}_u -0.0051 \tilde{d}_s) \ee
\end{itemize}
In this paper, we study the both formulae and discuss the effects of the
theoretical uncertainty.

 In the literature, to calculate the EDM of neutron various
theoretical approaches have been taken which give different and
even conflicting results. For example, the QCD sum rules yield
\cite{ritz} \be d_n=(1\pm 0.5) {|\langle \bar{q} q \rangle | \over
(225~{\rm MeV})^3} \times \left[ 0.55 e (\tilde{d}_d+0.5
\tilde{d}_u)+0.7(d_d-0.25 d_u) \right],\label{sumrule}\ee while
SU(3) chiral Lagrangian technique \cite{hisano} gives \be d_n
=(1.6 \times \tilde{d}_u+1.3 \times \tilde{d}_d +0.26 \times
\tilde{d}_s)\ \ {\rm e~cm}. \label{su(3)}\ee Notice that since we
expect $\tilde{d}_s/\tilde{d}_d \sim m_s/m_d\simeq 19$
\cite{mstomd}, in the latter formula the dominant contribution is
that of the strange quark whose effect is completely neglected in
the former formula. On the other hand, in the latter formula the
effects of EDM of quarks are neglected and only chromoelectric
dipole moments are considered. Because of these theoretical
uncertainties, we do not put much emphasis on the bounds from
neutron EDM. In our calculation, we will be using the formulation
in \cite{hisano}.

\begin{figure} \psfig{figure=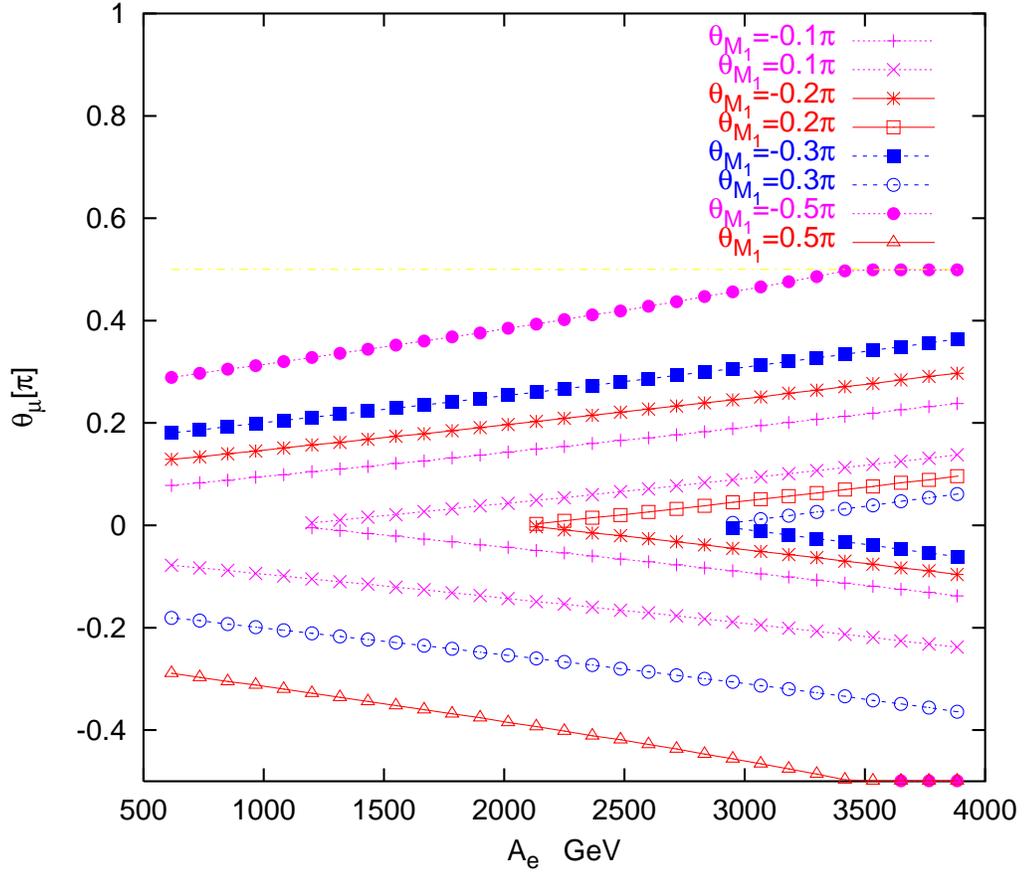,bb=33 37 553 501, clip=true,
height=5 in} \caption{Maximum  and minimum values of $\theta_\mu$
for which cancelation among the contributions of the phases of
$M_1$, $A_e$ and $\mu$ is possible. Each curve corresponds to a
different value of $\theta_{M_1}$ as indicated. We have taken
$m_{\tilde{e}_L}=190$~GeV, $m_{\tilde{e}_R}=125$~GeV,
$m_{\tilde{\nu}}=295$~GeV, $|M_1|=97$~GeV, $M_2=200$~GeV, $\tan
\beta=10$ which correspond to the values of these parameter for
the SPS1a' benchmark \cite{sps1a'}. However, the value of $|\mu|$
deviates from that of SPS1a': $|\mu|=440$~GeV.} \label{deM1Ae}
\end{figure}

Figs (\ref{deM2new},\ref{deM1Ae}) display the maximum values
of $\theta_\mu$ for which cancelation between the contributions of
the phases of $A_e$ and $\mu$ is possible. Since the masses of
selectrons and sneutrino are taken to be relatively small
($<1$~TeV), the dominant effects are given by one-loop chargino
and neutralino exchanges \cite{nath} and the effect of two-loop
diagrams can be neglected \cite{pilaftsis}.

 Drawing Fig.~\ref{deM2new}, the phases of gauginos are
 set equal to zero. From Fig.~\ref{deM2new} we conclude  that with
 growing
 $M_2$, cancelation for larger values of $\theta_\mu$ becomes
 possible. This is because the dominant contribution to $d_e$
 comes from the chargino exchange which depends only on the phase
 of $\mu$. That is while only the subdominant effect (the
 neutralino exchange diagram) depends on the phase of $A_e$.
 Increasing $M_2$, the effect of chargino exchange is suppressed
 and in turn cancelation between two effects will be possible for
 higher values of $\theta_\mu$.
 As expected, increasing the value of $|A_e|$, the range of values of $\theta_\mu$ for which cancelation
is possible is enlarged. This can be observed by comparing curves
(a1) with (a2); (a3) with (a4); and (b1) with (b2). Increasing
$\tan \beta$ from 10 to 20, the maximum values for which
cancelation is possible is considerably reduced [see curves (b1)
and (b2)]. Comparing curves (a1) and (a3) with (c1), (c2) and
(c3), we observe that by increasing the masses of the
supersymmetric particles the maximum $\theta_\mu$ for which
cancelation is possible is reduced. This rather counter-intuitive
behavior is also observed by \cite{plehn} for lower values of
$\tan \beta$ and $|A_e|$. However, as expected for larger values
of sfermion masses the degree of fine tuning of the phases
necessary for cancelation is smaller; {\it i.e.,} for larger
sfermion masses, the value of $
d_e^{exp}/d_e(\theta_\mu,\theta_{A_e}={\rm arbitrary })$ is
smaller. Finally comparing  curves (c1), (c2) and (c3) with each
other and contrasting (a1) and (a2) versus (a3) and (a4), we
observe that the results are robust against varying the values of
$|\mu|$.

Fig.~\ref{deM1Ae} shows maximum  and minimum values of
$\theta_\mu$ for which cancelation among the contributions of the
phases of $M_1$, $A_e$ and $\mu$ is possible. Each curve
corresponds to a different value of $\theta_{M_1}$ as indicated.
For input parameters, we have chosen the spectrum of the SPS1a'
benchmark which is compatible with all the present bounds and will
be investigated by the LHC \cite{sps1a'}. As far as the EDM bounds
are concerned, this benchmark is a typical of points close to it.
We confirmed the robustness of these results by varying the input
parameters around this point.
 When $\theta_\mu$ and
$\theta_{M_1}$ have opposite signs, cancelation becomes possible
for larger values of $|\theta_\mu|$ than in the case that  the
relative sign is positive. In the latter case, even values of
$|\theta_\mu|$ of order of one are compatible with the bounds on
$d_e$. Notice that the results are symmetric under $\theta_\mu \to
-\theta_\mu$ and $\theta_{M_1} \to -\theta_{M_1}$. If future
searches put stronger bounds on $d_e$, our results will still be
valid but a greater degree of fine tuning for successful
cancelation will be required.

\begin{figure} \psfig{figure=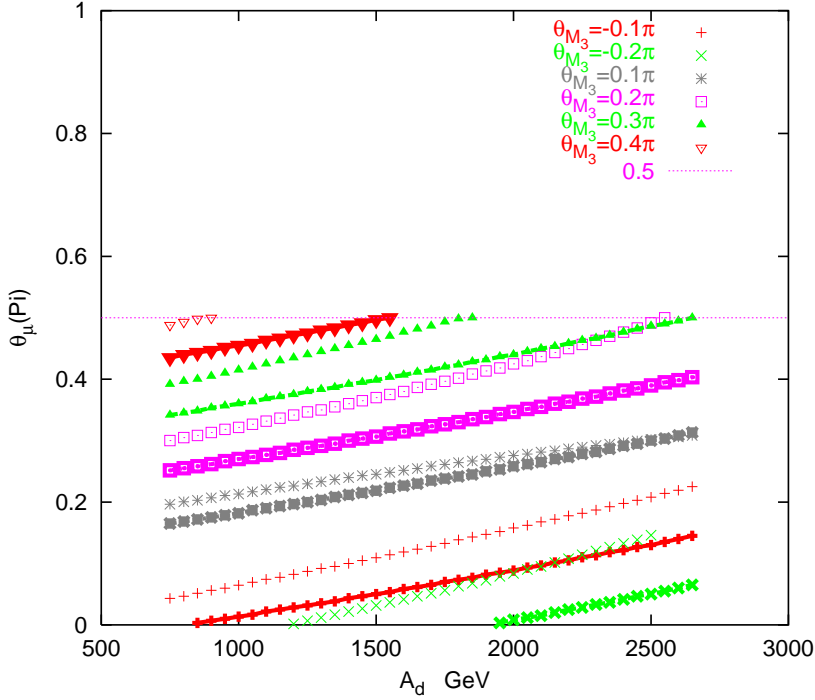,bb=33 37 553 501,
clip=true, height=4 in} \caption{Maximum values of $\theta_\mu$
for which cancelation among the contributions of the phases of
$A_d$, $A_s$, $M_3$ and $\mu$ to $d_{Hg}$ is possible. We have
taken $|A_u|=300$~GeV, $|A_s|=3000$~GeV, $m_d=6.3$~MeV,
$m_s=119$~MeV and $m_u=3$~MeV. Drawing the curves shown by
discrete thin points the formula (\ref{Hgfalk}) is used while for
the rest of the curves (thick points connected with lines) we have
used  Eq.~\ref{Hgshimizu}. We have taken $|\mu|=440$ GeV while the
rest of parameters correspond to the SPS1a' benchmark
\cite{sps1a'}: $m_{\tilde{d}_L}=m_{\tilde{s}_L}=565$~GeV,
$m_{\tilde{d}_R}=m_{\tilde{s}_R}=547$~GeV,  $M_1=97$~GeV,
$M_2=200$~GeV, $|M_{\tilde{g}}|=607$~GeV and $\tan
\beta=10$.}\label{HgM3Ad}
\end{figure}

 Let us now discuss the possibility of
cancelation between different contributions to $d_{Hg}$. This
possibility has been studied in Fig. (\ref{HgM3Ad}). Since, at the
electroweak scale, $m_{H_u}^2$ is negative (electroweak symmetry
breaking implies $m_{H_u}^2\simeq -|\mu|^2$), the CCB bounds
severely restrict the value of $|A_u|$. That is why we have taken
$|A_u|=300~{\rm GeV} \ll |A_d|,|A_s|$. For a given set of phases,
the contribution of $\tilde{d}_u$ to $d_{Hg}$ is subdominant. This
is expected because (i)  $|A_u| \ll |A_d|,|A_s|$; ii)
 the dominant contribution from $\theta_\mu$ to $\tilde{d}_u$
is suppressed by $\cot \beta$; that is while the corresponding
contribution to $\tilde{d}_d$ and $\tilde{d}_s$ is enhanced by
$\tan \beta$.
In general, we expect the uncertainty in the values of $m_d$ and
$m_s$ to dramatically affect the values of the calculated EDMs.
However, since the ratio $m_s/m_d$ is known to a rather high
accuracy \cite{mstomd} the region of parameter space in which
cancelation occurs is not considerably affected by the uncertainty
in knowledge of $m_d$. Nevertheless,   the theoretical uncertainty
in calculating $d_{Hg}$ [{\it e.g.,} Eq. (\ref{Hgshimizu}) vs. Eq.
(\ref{Hgfalk})] can change the conditions for cancelation to some
extent.

 Fig.~\ref{HgM3Ad}
displays the maximum values of $\theta_\mu$ for which cancelation
among the contributions of the phases of $M_3$, $\mu$, $A_d$,
$A_s$ and $A_u$ to $d_{Hg}$ is possible.   Each curve corresponds
to a different value of $\theta_{M_3}$. The same as in Fig.
\ref{deM1Ae}, we have set the input parameters close to the
benchmark SPS1a' \cite{sps1a'}. For a given value of
$\theta_{M_3}$, using the formula given in \cite{shimizu1} [see
Eq. (\ref{Hgshimizu})] the bound on $d_{Hg}$ appears more
restrictive than the formula given by the QCD sum rule \cite{falk}
[see Eq. (\ref{Hgfalk})]. This  can be observed from Fig.
\ref{HgM3Ad}: the thin discrete points lie above the thick points
connected by lines. This is expected because when the contribution
of $\tilde{d}_s$ is taken to be large, the phase of $A_s$ can play
a greater role in canceling  the other effects. From Fig.
\ref{HgM3Ad}, we conclude that the cancelation scenario is open
for a wider range of $\theta_\mu$ if the relative sign of
$\theta_{M_3}$ and $\theta_\mu$ is positive. This is because of
the partial cancelation that occurs between their respective
contributions in this case. When both $\theta_{M_3}$ and
$\theta_\mu$ are positive, even phases close to $\pi/2$ are
compatible with the bound on $d_{Hg}$.

\begin{figure}[h]
\begin{center}
\centerline{\vspace{-1.2cm}}
\centerline{\epsfig{figure=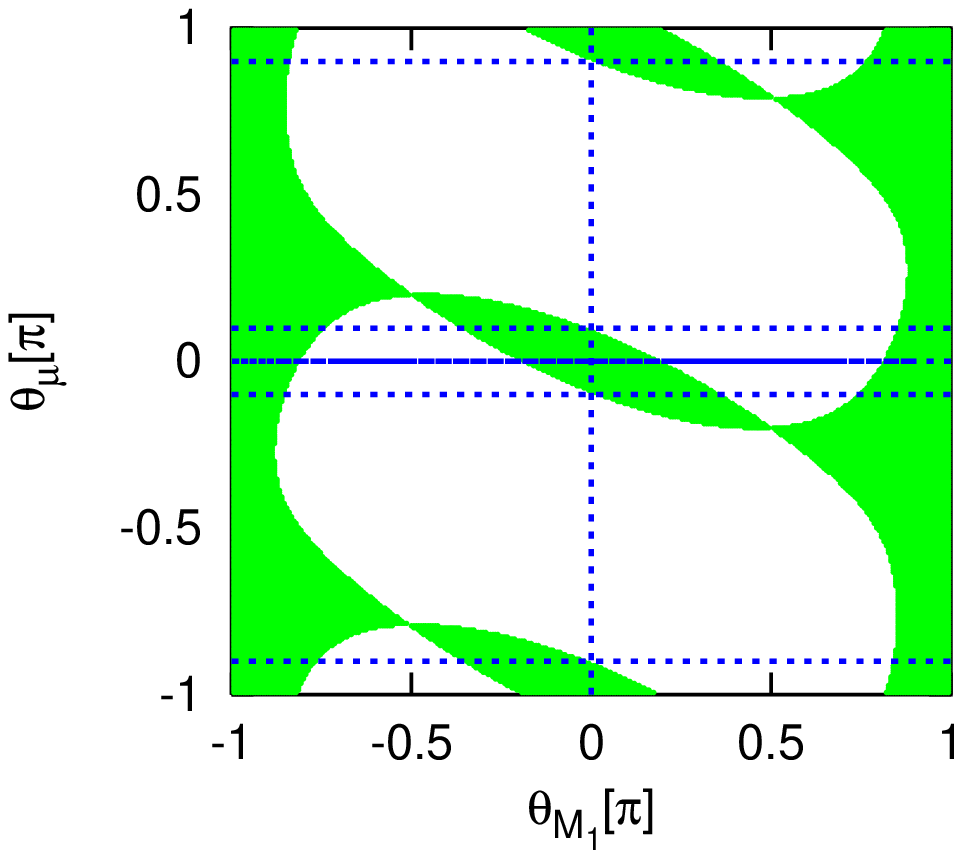,width=7cm}\hspace{5mm}\epsfig{figure=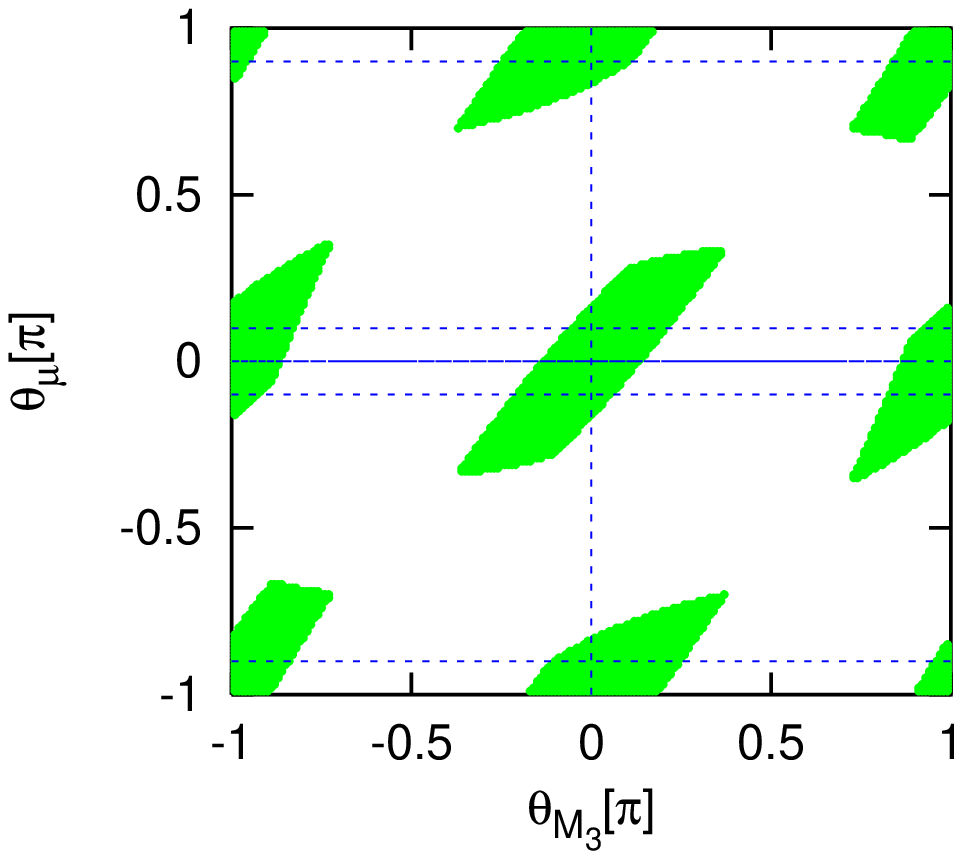,width=7cm}}
\centerline{\vspace{1.cm}\hspace{1cm}(a)\hspace{7cm}(b)}
\centerline{\vspace{-1.2cm}}
\centerline{\epsfig{figure=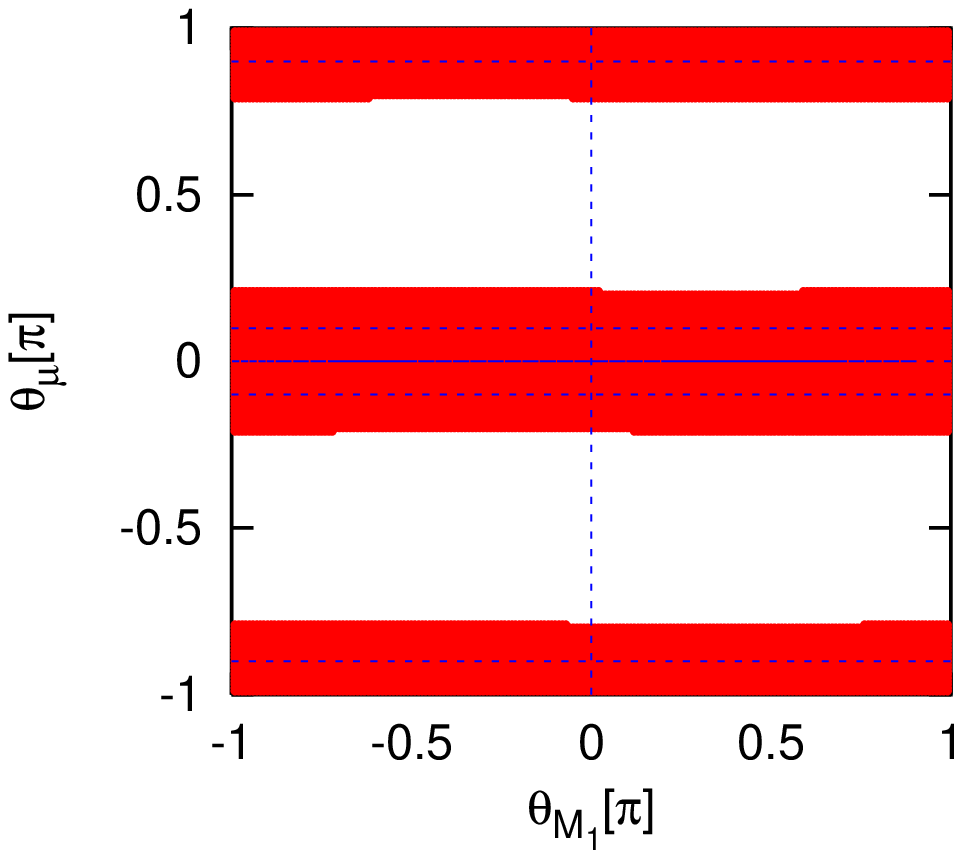,width=7cm}\hspace{5mm}\epsfig{figure=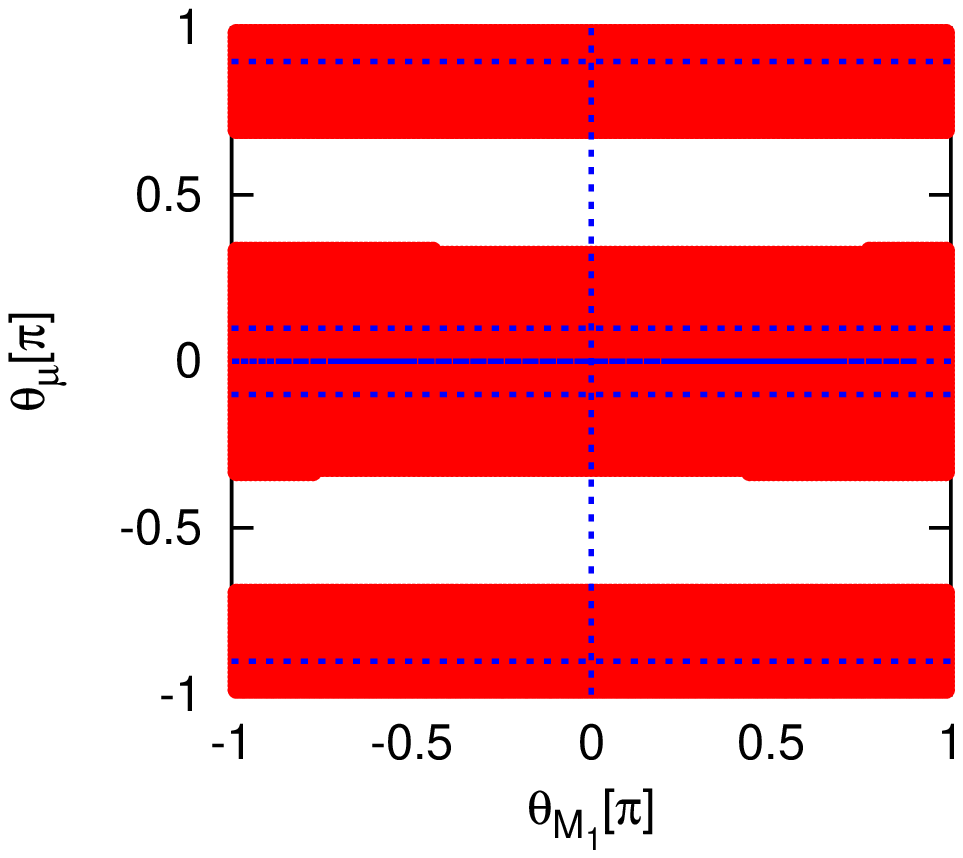,width=7cm}}
\centerline{\hspace{1.2cm}(c)\hspace{7cm}(d)}
\centerline{\vspace{-1.5cm}}
\end{center}
\caption{Shadowed areas depict the range of phases for which
cancelation is possible. We have set $|\mu|=400$ GeV and the
sfermion and gaugino masses correspond to their values for the
SPS1a' benchmark \cite{sps1a'} as explained in Figs.
(\ref{deM1Ae},\ref{HgM3Ad}). The horizontal dotted lines
correspond to $\theta_\mu=-0.9 \pi, -0.1 \pi, 0.1 \pi$ and $0.9
\pi$. a) The range of $\theta_\mu-\theta_{M_1}$ for which
cancelation between contributions of the phases of $A_e$, $\mu$,
and $M_1$ can yield zero $d_e$. We have set $|A_e|=2000$~GeV. b)
Region where cancelation between contributions of the phases of
the $A$-terms, $\mu$ and $M_3$ can yield  $d_n=d_{Hg}=0$. We have
set $\theta_{M_1}=0$ and taken $|A_u|=300$~GeV,
$|A_s|=|A_d|=3000$~GeV. c) The same as (b) except that here we
have set $\theta_{M_3}=0$. d) The same as (b,c) except that here
all phases ($\theta_{M_1}$, $\theta_{M_3}$, $\theta_\mu$,
$\theta_{A_u}$, $\theta_{A_d}$ and $\theta_{A_s}$) are allowed to
vary. }\label{overlap}
\end{figure}

Remember that
 we have allowed the $A$-parameters of the electron and quarks to
be different from each other, so the degrees of freedom are enough
to simultaneously fulfill the cancelation conditions for $d_e$ and
$d_{Hg}$  with non-vanishing solutions for phases. However one
should check if there is an overlap between the range of
$\theta_\mu$ and $\theta_{M_1}$ for which $d_e$ and $d_{Hg}$ can
vanish  because of cancelation. Relaxing the assumption of
universality, this comparison is rather non-trivial because
$d_{Hg}$ and $d_e$ are sensitive to different sets of input
parameters. Since the mass spectrum for both figures \ref{deM1Ae}
and \ref{HgM3Ad} correspond to the same benchmark, it makes sense
to compare them. To make such a comparison, we should notice that
1) $d_e$ does not depend on $\theta_{M_3}$; 2) sensitivity of
$d_{Hg}$ to $\theta_{M_1}$ is low. Comparison shows that we can
simultaneously satisfy both bounds  for $|\theta_\mu|\sim {\cal
O}(1)$. Figs. \ref{overlap} also show the range of
($\theta_\mu,\theta_{M_1}$) [or $\theta_\mu,\theta_{M_3}$] for
which cancelation is possible. To draw these figures, we have
taken the spectrum of SPS1a' as input for masses but we have taken
$|A_e|=2000$~GeV, $|A_u|=300$~GeV and $|A_s|=|A_d|=3000$~GeV. Fig
(\ref{overlap}-a) shows the range of $\theta_\mu$ and
$\theta_{M_1}$ for which cancelation between contributions of
phases of $\mu$, $M_1$ and $A_e$ to $d_e$ is possible. Notice that
even $\theta_\mu=\pi/2$ can be compatible with the upper bound on
$d_e$. In the area shadowed in Fig. (\ref{overlap}-b), cancelation
between contributions of the phases of $\mu$, $M_3$, $A_u$, $A_d$
and $A_s$ to $d_n$ and $d_{Hg}$ can suppress   $d_n$ and $d_{Hg}$
to  values below  the strong bounds on them. To draw Fig.
(\ref{overlap}-b), we have set $\theta_{M_1}=0$. Fig.
(\ref{overlap}-c) is similar to Fig. (\ref{overlap}-b) except that
in the case of Fig. (\ref{overlap}-c), $\theta_{M_3}$ instead of
$\theta_{M_1}$ is set equal to zero.
 Fig. (\ref{overlap}-d)  displays the range of $\theta_\mu$ and $\theta_{M_1}$ for which
cancelation between different contributions to $d_n$ and $d_{Hg}$
is possible. Drawing Fig (\ref{overlap}-d) the phases
$\theta_\mu$, $\theta_{M_1}$, $\theta_{M_3}$, $ \theta_{A_d}$,
$\theta_{A_u}$ and $\theta_{A_s}$ are all allowed to vary in order
to satisfy the bounds on $d_n$ and $d_{Hg}$. As expected, Figs.
(\ref{overlap}-c) and (\ref{overlap}-d) look  similar but Fig.
(\ref{overlap}-d) covers a larger range because in this case there
is one more degree of freedom to satisfy the bounds. To
simultaneously satisfy all the bounds, the values of  $\theta_\mu$
and $\theta_{M_1}$ should be in the overlap of Figs
(\ref{overlap}-a) and (\ref{overlap}-d). Contrasting these two
figures we find out that values of $\theta_\mu$ and $\theta_{M_1}$
of order of 1 are possible. It is interesting that the maximal
CP-violating case $\theta_\mu=\pm \frac{\pi}{2}$ which is
compatible with $d_e$ is excluded by the bounds on $d_{Hg}$ and
$d_n$ and on the other hand, $\theta_{M_1}=\pm \frac{\pi}{2}$
which is compatible with the $d_n$ and $d_{Hg}$ bounds is ruled
out by bound on $d_{e}$ . One should be aware that in this range
of parameters the bound from $d_n$ is more restrictive than the
bound from $d_{Hg}$. If we redraw the Figs. (\ref{overlap}-b) to
(\ref{overlap}-d) overlooking the bound on $d_{Hg}$, we find that
results do not considerably change. This shows that in order to
derive conclusive results from EDMs, solving the theoretical
uncertainties in calculation of $d_n$ is imperative.

\begin{figure}[h]
\begin{center}
\centerline{\epsfig{figure=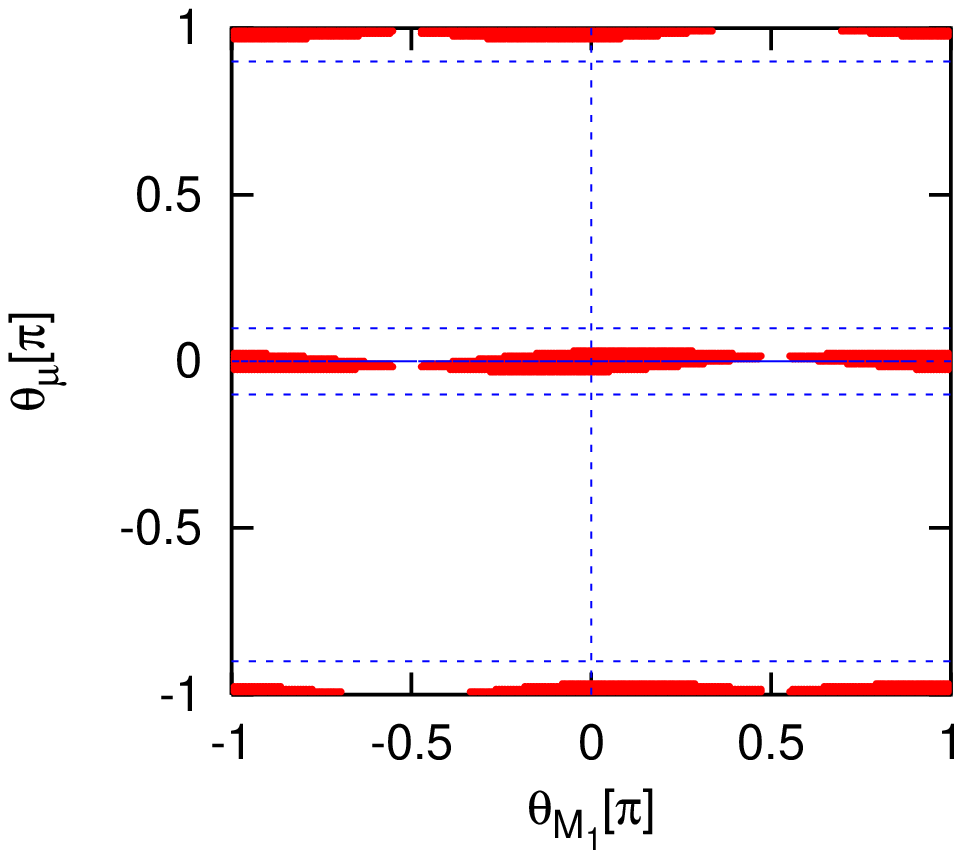,width=7cm}\hspace{5mm}\epsfig{figure=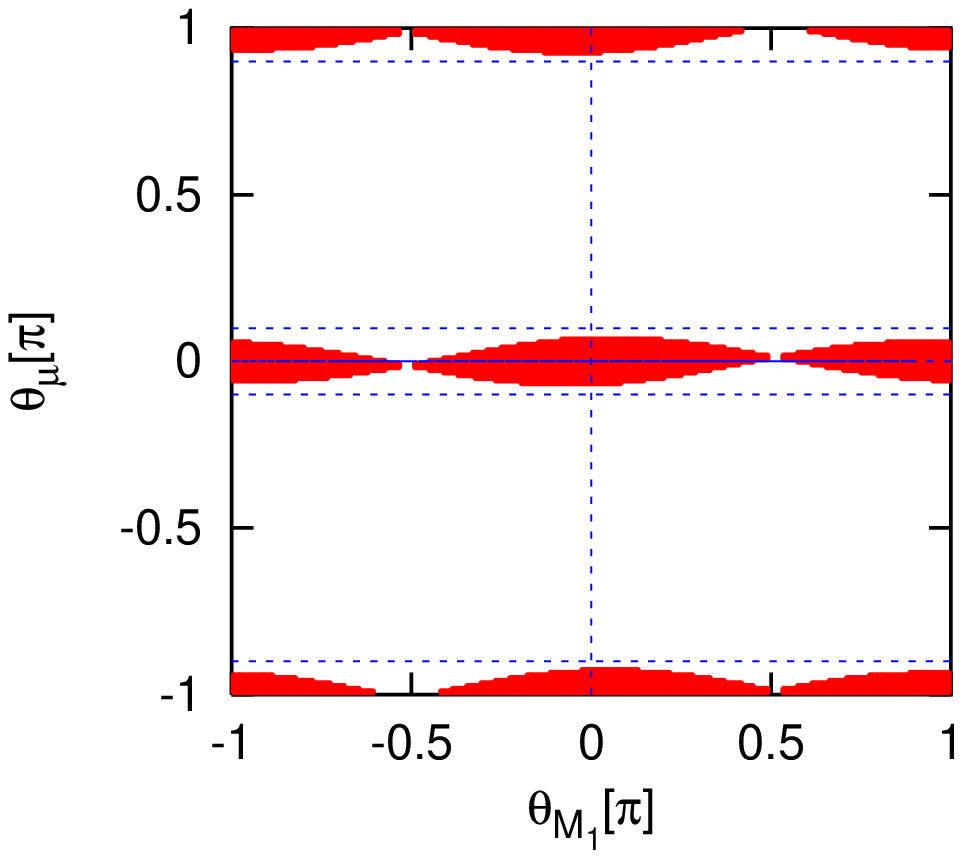,width=7cm}}
\centerline{\vspace{1.cm}\hspace{1cm}(a)\hspace{7cm}(b)}
\end{center}
\caption{The shadowed areas show the values of phases for which
different contributions to $d_e$ can cancel each other. The
horizontal dotted lines correspond to $\theta_\mu=-0.9 \pi, -0.1
\pi, 0.1 \pi$ and $0.9 \pi$. We have chosen input parameters such
that  the Bino-Higgsino mixing is sizeable: a) $|\mu|=200$~GeV,
$|M_1|=150$~GeV, $M_2=312$~GeV, $m_{\tilde{e}_L}=336$~GeV,
$m_{\tilde{e}_R}=223$~GeV, $m_{\tilde{\nu}_L}=327$~GeV,
 $|A_e|=2000$~GeV and $\tan \beta=10$;
b) the same as (a) except $|A_e|=4000$~GeV. } \label{HBmixing}
\end{figure}

From cosmological point of view, the region $\mu\simeq M_1$ is
specially interesting because in this region, the Higgsino Bino
mixing is sizeable yielding relatively large
$\tilde{\chi}_1^0\tilde{\chi}_1^0$ annihilation rate in accord
with the dark matter density measurements. Recently it has been
shown \cite{semenov}  that varying the values of the CP-violating
phases, the range of parameters compatible with the measured dark
matter density will be enlarged. Fig.~\ref{HBmixing} tries to
find out if large CP-violating phases, for sub-TeV selectron
masses, are allowed. From these figures, we find out that although
$\theta_{M_1}$ can be ${\cal O}(\pi/2)$,  $\theta_\mu$ cannot
exceed 0.1$\pi$ even if $A_e=4000$~GeV. We have checked the
robustness of this result by varying the parameters along
$\mu\simeq M_1$ and the result seems generic.

 \section{EDM bounds and electroweak baryogenesis}

\begin{figure} \psfig{figure=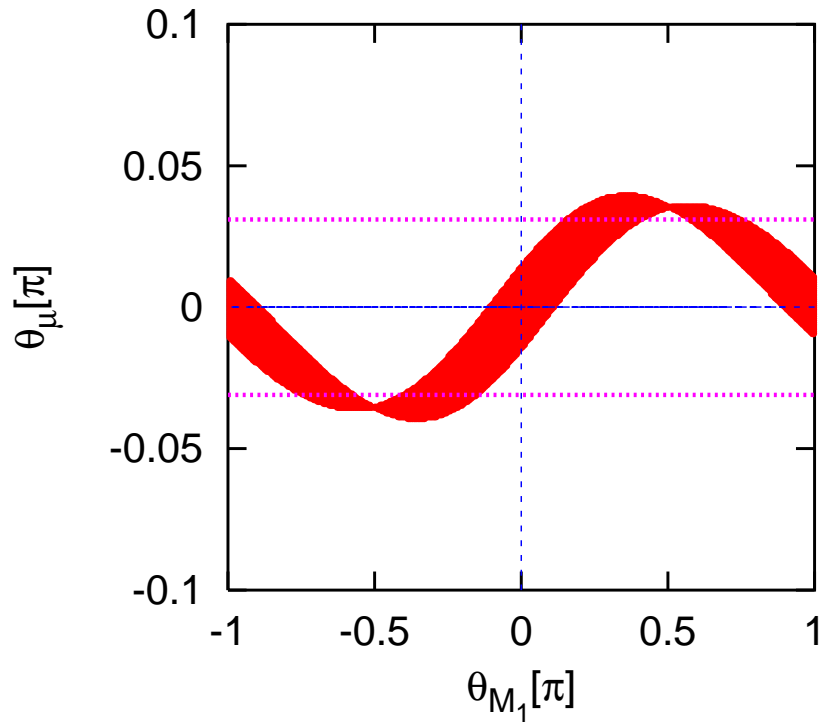, bb=50 38 340 297,
clip=true, height=4 in} \caption{The range of phases of $\mu$ and
$M_1$ for which total cancelation among the contributions of the
phases $\mu$, $M_1$ and $A_e$ to $d_e$ is possible. We have taken
$m_{\tilde{e}_L}=392$~GeV, $m_{\tilde{e}_R}=220$~GeV,
$m_{\tilde{\nu}_L}=385$~GeV, $|A_e|=700$~GeV, $|M_1|=200$~GeV and
$M_2=415$~GeV and $\tan \beta=10$. We have set $|\mu|=200$~GeV$=
|M_1|$ which corresponds to the neutralino-driven resonance
condition of electroweak baryogenesis. The purple dotted
horizontal lines depict $\sin \theta_\mu=\pm 0.1$.}
\label{baryogenesis200}
\end{figure}

\begin{figure} \psfig{figure=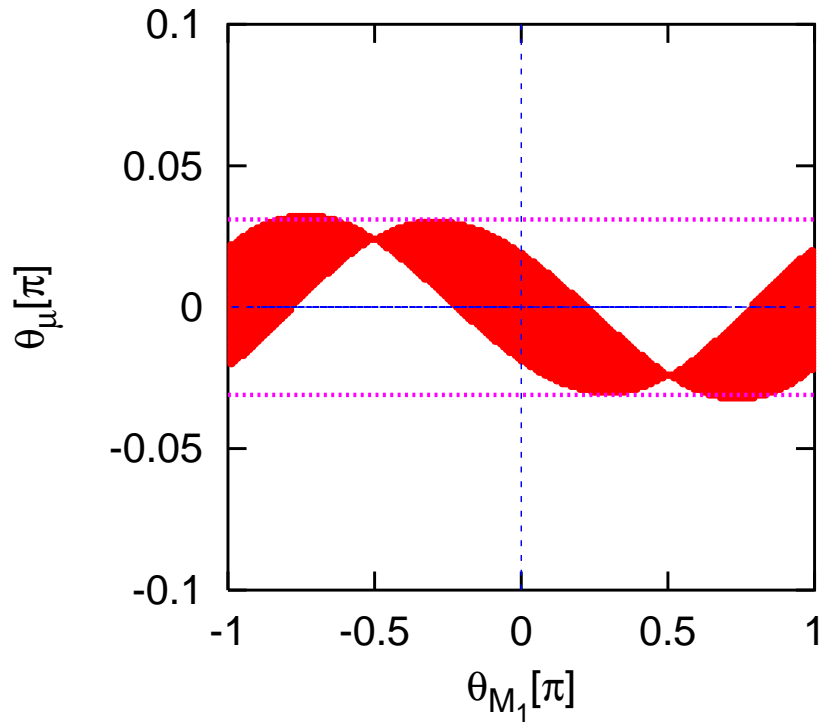, bb=50 38 340 297,
clip=true, height=4 in} \caption{The range of phases of $\mu$ and
$M_1$ for which total cancelation among the contributions of the
phases $\mu$, $M_1$ and $ A_e$ to $d_e$ is possible. We have taken
$m_{\tilde{e}_L}=333$~GeV, $m_{\tilde{e}_R}=187$~GeV,
$m_{\tilde{\nu}_L}=324$~GeV, $|A_e|=700$~GeV, $|M_1|=167$~GeV and
$M_2=348$~GeV and $\tan \beta=10$. We have set
$|\mu|=340$~GeV$\simeq M_2$ which corresponds to the
chargino-driven resonance condition of electroweak baryogenesis.
The purple dotted horizontal lines depict $\sin \theta_\mu=\pm
0.1$.}\label{baryogenesis340}
\end{figure}

Arguably the most important manifestation of   CP-violation  is
its role in the creation of the baryon asymmetry of the Universe.
In the context of MSSM, all three Sakharov's famous conditions can
be fulfilled and, in principle, through a mechanism known as
electroweak baryogenesis, the baryon asymmetry of the Universe can
be explained. To have strong first order electroweak phase
transition necessary for the creation of the baryon asymmetry of
the universe, one of the top squarks should be lighter than the
top quark. If we demand the lightest neutralino to be the lightest
supersymmetric particle, this in turn implies
$m_{\tilde{\chi}^0_1}<m_t$. The other requirement for having
successful electroweak baryogenesis is of course having large
enough CP-violating phases. However, in \cite{stefano} it is shown
that even  for values of $\sin\theta_\mu$ as low as $10^{-2}$
successful electroweak baryogenesis can be a possibility provided
that we are at the resonance region \cite{ramsey,resonance} (i.
e., $|\mu|\simeq |M_1|$ or $|\mu|\simeq |M_2|$). Moreover to have
a successful electroweak baryogenesis the mass of the CP-odd Higgs
boson, $m_{A^0}$, should be relatively low ($m_{A^0}\ll $~TeV).
Notice that if the masses of selectron and sneutrino are below the
TeV scale, even values of $\sin\theta_\mu$ as low as $10^{-2}$
will not be compatible with the bounds on the electric dipole
moment of electron, unless the cancelation scenario is at work.
Suppose future experiments (the LHC and ILC) confirm supersymmetry
and find out that $m_{\tilde{\chi}_1^0}<m_{\tilde{t}_R}<m_t$ and
discover a relatively light $A^0$. These conditions are
tantalizingly close to the requirement for a successful
electroweak baryogenesis. Now, suppose that the masses of
selectrons turn out to be at the scale of few hundred GeV. Does
this mean that the electroweak baryogenesis is ruled out? Figs
(\ref{baryogenesis200},\ref{baryogenesis340}) try to address this
question by studying the possibility of cancelation between
different contributions to $d_e$. As we discussed in the previous
section, since we have set the masses of sfermions below the TeV
scale, the one-loop effects dominate over the two-loop effects.
Taking the two-loop effects into account only slightly shifts the
cancelation point.

 To draw Fig \ref{baryogenesis200}, we have set $\tan \beta
=10$, $m_{\tilde{e}_L}=392$~GeV, $m_{\tilde{e}_R}=218$~GeV,
$m_{\tilde{\nu}_L}=385$~GeV, and $M_2=415$~GeV. Moreover we have
set  $|M_1|=|\mu|=200$ GeV which means we are in the
neutralino-driven resonant electroweak baryogenesis regime
\cite{stefano}. For this choice of parameters the lightest
neutralino is indeed lighter than the top quark. We have set
$A_e=700$~GeV which is smaller than
$[3(m_{\tilde{e}_L}^2+m_{\tilde{e}_R}^2)]^{1/2}$ thus,  as long as
$m_{H_d}^2$ is positive \cite{gunion}, no CCB will take place (see
Eq.~\ref{aebound}). Positiveness of $m_{H_d}^2$ sets a lower bound
on $b\tan\beta\simeq m_{A^0}^2$ [see Eq. (\ref{Mhd})] which for
our choice of parameters is 190~GeV. Thus, for these parameters
$A^0$ (the CP-odd Higgs boson) can  still be sufficiently light.
Increasing $A_e$ the cancelation can of course become possible for
larger values of $\theta_\mu$ but the lower bound on $m_{A^0}$
will  also be stronger and on the other hand, for heavier
$m_{A^0}$ the produced baryon asymmetry is suppressed. As shown in
\cite{stefano}, the neutralino-driven resonant baryogenesis is
only marginally compatible with the indirect searches of dark
matter so this choice of parameters in near future will be tested
not only by collider data but also by further indirect dark matter
searches.

From Fig. \ref{baryogenesis200}, we observe that for universal
gaugino masses [$\theta_{M_1}=\theta_{M_2}=0$], cancelation can
take place even for values of  $|\sin \theta_\mu|$ up to  0.06
which according to \cite{stefano} can easily yield the
baryon-antibaryon asymmetry compatible with the WMAP results. This
confirms the results of \cite{ramsey}. In the neutralino-driven
electroweak baryogenesis regime, the combination of the phases
which determines baryogenesis is $\theta_\mu+\theta_{M_1}$
\cite{private}. Notice that $\theta_\mu+\theta_{M_1}$ is a
rephasing invariant quantity. Fig \ref{baryogenesis200}  shows
that, relaxing the assumption of the universality of the gaugino
masses [$\theta_{M_1}\ne \theta_{M_2}=0$], cancelation makes
$|\sin(\theta_\mu +\theta_{M_1})|\sim 1$ compatible with the
bounds on $d_e$.

Now let us discuss fine tuning required for such cancelation.  If
the phases of $\mu$ and $M_1$ are at the region where cancelation
can take place, the generic value of $d_e$ is already around
$10^{-26}$~e cm so to reduce the value of $d_e$ down to below the
upper bound on it (see Eq.~\ref{10-27}), a cancelation of $10\%$
will be enough which means the fine tuning of the phases is not a
problem. It can be shown that for this range of $\theta_\mu$ and
$\theta_{M_1}$, the different contributions from phases of $M_3$,
$M_1$, $\mu$, $A_d$ and $A_s$
 to $d_n$ and $d_{Hg}$ can also cancel each other to satisfy the
experimental upper bounds. The degree of fine-tuning necessary for
suppressing the values of  $d_n$ and $d_{Hg}$ down to the values
below the upper bounds on them depends on the scale of $M_3$ and
squark masses as well as $A$-terms. We have checked the degree of
fine tuning needed for effective cancelation setting $M_3\simeq
1400$~GeV and $m_{\tilde{d}}=1200$~GeV (these are the values
corresponding to the prediction of cMSSM if the values of $M_1$
and selectron are as above). We have  set $A_s=A_d=900$~GeV which
are smaller than $\sqrt{3(m_{\tilde{d}_L}^2+m_{\tilde{d}_R}^2)}$.
For  the phases in the shadowed region shown in Fig
\ref{baryogenesis200} cancelation required in order to suppress
$d_{Hg}$ below the upper bound on it is not stronger than $\sim$
5\% ($\sim$ 20\%) if we take the formula in \cite{falk}
(\cite{shimizu1}). We have also found that cancelation required in
order to suppress $d_{n}$ below its upper bound is not stronger
than $\sim$ 3\% ($\sim$ 15\%) if we take the  formula in
\cite{hisano} (\cite{ritz}). In near future, the experiments are
going to become  sensitive to even smaller values of $d_n$,
$d_{Hg}$ and $d_{e}$. Moreover, there are proposals to probe EDM
of deuteron down to $(1-3)\times 10^{-27}$~e~cm \cite{dD}. If one
or more of these experiments detect a nonzero electric dipole
moment, it will be a strong hint in favor of  the electroweak
baryogenesis. On the other hand, if they all report null results,
we still  cannot rule out the cancelation scenario even though a
new piece of information (the bound on $d_D$) is added. However a
greater degree of fine tuning would be necessary for the
cancelation.

Fig. \ref{baryogenesis340} explores the possibility of cancelation
scenario and having large CP-violating phases in the
chargino-driven resonant electroweak baryogenesis regime
($|\mu|\simeq M_2$). The above discussion holds in this case, too,
with the difference that for the chargino-driven electroweak
baryogenesis the combination of phases that relevant for
baryogenesis is $\theta_\mu+\theta_{M_2}$. According to this
figure $\sin(\theta_\mu+\theta_{M_1})$  can reach 0.09 which may
be enough for a successful baryogenesis \cite{stefano}. In case of
Fig. \ref{baryogenesis340}, since $|\mu|$ is larger, the lower
bound on the CP-odd Higgs boson will be stronger:
$m_{A^0}>335$~GeV. Unlike the case of neutralino-driven
electroweak baryogenesis,
 the chargino-driven electroweak baryogenesis is not sensitive to the indirect dark matter
searches.


\section{Implication of cancelation scenario for CP-violation searches in the colliders} 
\begin{figure}[h]
\psfig{figure=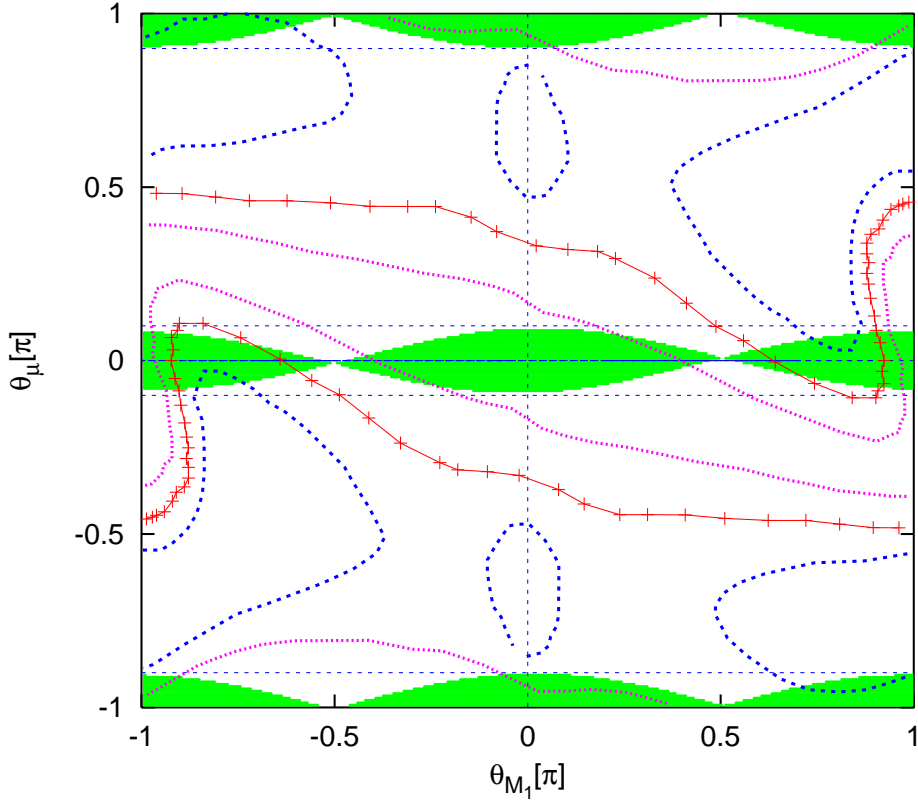,bb= 33 37 553 501, clip=true,
height=5 in} \caption{Shadowed areas show the region where cancelation can yield vanishing
$d_e$. The curves, which are borrowed from Fig. 2.12.b of \cite{kittelthesis},
 correspond to various values of $A_{CP}$ (see the text for  the definition of $A_{CP}$):
 Dashed lines correspond to $A_{CP}=\pm 45$\%; curves marked with $+$ indicate $A_{CP}=\pm 30$\%
 and finally the pink curves correspond to $A_{CP}=\pm 15$ \%. To draw the shadowed area we have
 used the same input parameters as in Fig 2.12.b of \cite{kittelthesis}:
 $|\mu|=300$~GeV,
$m_{\tilde{e}_L}=378$~GeV, $m_{\tilde{e}_R}=211$~GeV,
$m_{\tilde{\nu}_L}=370$~GeV, $|M_1|=192$~GeV, $M_2=400$~GeV,
$|A_e|=2000$~GeV and $\tan \beta=5$. The horizontal dotted lines
correspond to $\theta_\mu=-0.9 \pi, -0.1 \pi, 0.1 \pi$ and $0.9
\pi$. } \label{figkittel}
\end{figure}

The
CP-violating phases can manifest themselves as both CP-even and CP-odd quantities in the LHC
\cite{godbole,cplhc} and International Linear Collider,  ILC \cite{cpilc,kittelthesis,kittel}.
Due to high precision and capability of polarizing the initial beams, the ILC will have a
greater chance to observe CP-violation in the production and decay of sparticles.
In \cite{kittelthesis,kittel}, it is shown that even small values of CP-violating phases can
result in an asymmetry between $e^+e^- \to \tilde{\chi}_1^0 \tilde{\tau}_1^+ \tau^-$
and $e^+e^- \to\tilde{\chi}_1^0 \tilde{\tau}_1^- \tau^+$. Following \cite{kittelthesis},
let us define
\begin{equation} A_{CP}\equiv {P_2-\bar{P}_2 \over 2}. \end{equation}
In the above definition, $P_2$ is the polarization of $\tau$ which is produced in the subsequent processes
$e^+ e^- \to \tilde{\chi}_1^0 \tilde{\chi}_i^0$ and $\tilde{\chi}_i^0 \to \tau^-\tilde{\tau}^+$.
The polarization vector is
defined as
\begin{equation}
\vec{P}\equiv \frac{{\rm Tr}[ \rho \vec{\sigma}]}{{\rm Tr}[\rho]},
\end{equation}
where $\rho$ is the spin density of $\tau$ and direction 2 is
taken to be perpendicular to the plane defined by the momenta of
the $\tau$ and the initial electron.  Curves in Fig.
\ref{figkittel}, which are borrowed  from Fig. 2.12.b of
\cite{kittelthesis}, show different contour lines corresponding to
various values of $A_{CP}$. The input data for the curves are
$\theta_{A_\tau}=0$, $A_\tau=250$~GeV and
($P_{e^-},P_{e^+})=(-0.8,0.6)$. The rest of the input parameters
are given in the caption of Fig. \ref{figkittel}. Notice that  the
input parameters satisfy the relations that we would have expected
in the cMSSM. It is remarkable that  $A_{CP}=\pm 45\% $ can be
possible for values of $\theta_\mu$ as small as $\pm 0.1 \pi$ and
$\theta_{M_1}=\pm 1/6\pi$ or for $\theta_\mu=\pm 0.06 \pi$ and
$\theta_{M_1}=\pm \pi/2$. The shadowed areas superimposed on the
curves show the region for  which the cancelation scenario can
result in vanishing $d_e$. In order to check if in the same area
vanishing $d_n$ and $d_{Hg}$ is possible, we  calculated the
corresponding gluino and squark masses in the specific point in
the cMSSM space chosen above and inserted them in the formulae for
$d_{Hg}$ and $d_n$. We found that for any given set of
$\theta_\mu$ and $\theta_{M_1}$ total cancelation can
simultaneously suppress the values of $d_n$ and $d_{Hg}$. The
overlap of curves with the shadowed area indicates that even for
light sfermion masses, we still have a hope to observe
CP-violating effects in the ILC provided that the systematic and
statistical errors are under control.

Let us now discuss the fine tuning required for suppressing the
EDMs below the upper bounds on them. Taking $\theta_\mu$ and
$\theta_{M_1}$ in  the shadowed area and assigning a general value
between $-\pi$ and $\pi$ to $\theta_{A_e}$ we find that $d_e$
cannot exceed $10^{-26}$~e cm. This means that the fine tuning
required to suppress $d_e$ below the bound in Eq. \ref{10-27} is
not larger than 10\%. However, although simultaneous suppression
of $d_n$ and $d_{Hg}$ is possible for a wider range of phases, we
have found that the required fine-tuning in this case is greater
and is of order of 1\%.
\section{Conclusions}
In this paper, we have studied the possibility of satisfying the
bounds on $d_e$, $d_{Hg}$ and $d_n$ by cancelation scenario,
relaxing the universality of parameters at the GUT scale. We have
discussed that  relaxing universality of the Higgs mass parameters
($m_{H_d}^2$ and $m_{H_u}^2$) the color and charge breaking bounds
on $A_e$, $A_d$ and $A_s$ will be less severe allowing $A$-terms
of order of few TeV. We have focused on the part of parameter
space with intermediate values of $\tan \beta$ ($\tan \beta \simeq
10$) and TeV scale A-terms. This range has not been explored in
the literature before. We have found that the bounds on $d_e$ and
$d_n$, in this range, are complementary. We have argued that the
uncertainty in evaluating $d_n$ in terms of the EDMs and CEDMs of
the elementary particles causes large uncertainty in deriving
bounds on the CP-violating phases. For the mass spectrum close to
that of the SPS1a' benchmark (which will be soon explored at the
LHC) $|\sin \theta_\mu|,|\sin \theta_{M_1}|\sim {\cal O}(1)$ is a
possibility but requires high fine tuning (0.1\%).

 We have also
studied the possibility of cancelation for the region that
electroweak baryogenesis is enhanced ($|\mu| \simeq |M_1|$ and
$|\mu|\simeq |M_2|$) and found that $|\sin \theta_\mu|\simeq 0.1$
and $|\sin \theta_{M_1}| \simeq 1$, even for the sub-TeV slepton
masses, can be compatible with the EDM bounds. The main point is
that relaxing the assumption of the universality of gaugino mass
phases $(\theta_{M_1}\ne \theta_{M_2})$ opens up the possibility
of cancelation such that values of $\left| \sin [\theta_\mu
+\theta_{M_1}]\right| \sim 1$ become compatible with the bounds on
$d_e$. This opens new windows towards successful electroweak
baryogenesis.  Notice that in this range of parameters the
fine-tuning required for successful cancelation is not too high.

 We have explored the possibility of
cancelation in the regime that the Higgino-Bino mixing is large.
This part of the parameter space is of interest from neutralino
relic density point of view.
 In the literature \cite{semenov}, it is shown that varying the CP-violating phases in a large range broadens
the parameter range compatible with the WMAP dark matter
measurements. We find that in this regime, $|\sin\theta_\mu|$
cannot exceed 0.1 but $|\sin \theta_{M1}|$ can reach 1, even for
relatively small sfermion masses.

We expect in the future, the running and planned experiments to
probe smaller values of $d_n$ and $d_{Hg}$ and $d_e$
\cite{prospect}. Moreover the proposed experiments can probe $d_D$
with high sensitivity \cite{dD}. Even if all these experiments
report null results, the range of parameters where cancelation is
possible will not change but greater degree of fine tuning will be
required.

\section{Acknowledgement}
We would like to thank D. Demir, J. Ellis, R. Godbole and  M.
Hashemi for the useful discussions. We are specially thankful to
S. Profumo for pointing out a mistake in the first version of the
manuscript. We appreciate O. Kittel for giving us the  permission
to use his plots. We are also grateful to R. Asgari for teaching
us how to use the software for drawing the plots. Y. F. is
grateful to the CERN theory division for the hospitality during
her stay at CERN when a part of this work was done.


\end{document}